\documentclass[aps,groupedaddress,showpacs]{revtex4}
\usepackage[dvips]{graphicx}
\usepackage[]{caption}
\usepackage{amsmath}
\usepackage{amssymb}
\def\Dsl{\hbox{/\kern-.6700em\it D}} 
\def\dsl{\hbox{/\kern-.5300em$\partial$}}

\def\beq{\begin{equation}}
\def\eeq{\end{equation}}

\def\beqan{\begin{eqnarray*}}
\def\eeqan{\end{eqnarray*}}
\def\beqa{\begin{eqnarray}}
\def\eeqa{\end{eqnarray}}

\begin{document}

\bibliographystyle{prsty}
\title{Concern Regarding ``A Non-Inflationary Solution to the Entropy Problem of Standard Cosmology''.}
\author{Natalia Shuhmaher}
\email[Email:]{natalia.shuhmaher@physics.unige.ch}
\affiliation{D\'{e}partement de Physique Th\'{e}orique,
Universit\'{e} de Gen\'{e}ve, 24 Ernest Ansermet, 1211 Gen\'{e}ve 4,
Switzerland}
\pacs{98.80.Cq}
\begin{abstract}

We discuss the entropy and the size/homogeneity/horizon problems in
power-law expanding universes with one scale initial conditions. We
set the minimal scale $\Lambda = 10^{25}$ GeV at which a
non-inflationary solution is possible and show that the radiation
dominated epoch alone technically is not able to explain the issue.
An earlier paper on the subject,~\cite{size}, attempts a
multidimensional scenario. We review the scenario of the
paper~\cite{size} in an effective four dimensional Einstein frame.
We find that during contraction of extra dimensions, the residual
bulk energy density is important and leads to a scaling solution.
The existence of this scaling solution is a new result. We provide a
numerical example which demonstrates the evolution of the scale
factor and the extra dimensions. In the whole, the validity of the
effective field theory calculations at the scale of $\Lambda =
10^{25}$ GeV is under question and, hence, the final conclusion
regarding the possibility of a non-inflationary solution is
preliminary.

\end{abstract}

\maketitle

\section{Introduction}

One of the puzzles of the Big Bang Cosmology is the entropy problem.
During the adiabatic expansion of the universe, the entropy per
co-moving volume ($S$) is constant, and \beq S \propto g_{*}
\alpha^3 T^3 \, , \eeq where $g_{*}$ is the number of
ultra-relativistic degrees of freedom, $\alpha$ is the scale factor
and $T$ is the temperature. $g_{*}$ does not change by more than a
couple orders of magnitude during the history of the universe and we
neglect its time dependence in the following analysis. The entropy
of the volume corresponding to the current Hubble patch, $H^{-1}_0
\approx 10^{42} \, GeV^{-1}$~\cite{Hubble:1929ig,Freedman:2000cf},
which is stored in relativistic degrees of freedom is \beq S_U =
\frac{8 \pi^3}{135} g_{*} H^{-3}_0 T^3_0 \simeq 10^{88} \, . \eeq
where $0$ stands for current values. The value of $S_U$ represents a
puzzle. It takes its roots in the value of the present Hubble
parameter, $H_0$. The problem is called the entropy problem.

This puzzle together with the homogeneity of the observed
universe~\cite{Hinshaw:2008kr} set the size problem. Going back to
the Planck epoch one discovers that universe grew by a factor of
\beq \frac{T_p}{T_0} = 10^{32} \, , \eeq  where $T_p \, = \, M_p \,
= \, 1.22 \, \, { \mbox x } \, 10^{19} GeV$ is the Planck scale (in
the following, p stands for Planck epoch values). The physical size
at the Planck epoch corresponding to the current Hubble radius
($H_0^{-1}$) is $l_{H_0} = 10^{10} \, GeV^{-1}$. In a power law
expanding Universe the Hubble radius is roughly equal to the maximal
causally connected region (horizon). However, at the Planck epoch
the size of the universe corresponding to the Hubble radius at that
time is $l_p=10^{-19} \, GeV^{-1}$. Hence, $10^{87}$ causally
disconnected patches had to have approximately the same initial
conditions in order that we observe homogeneous universe
today~\cite{Hinshaw:2008kr}. This is the size problem which
sometimes is called the homogeneity~\cite{PFCMukhanov} or
horizon~\cite{KolbTurn} problem. Note that the number of patches
($N$) is \beq N = \frac{E_p}{M_p} = \frac{3}{4} \, S_U \eeq where
$E=3/4 \, S_U T$ is the energy of the relativistic degrees of
freedom in the volume corresponding to the current Hubble patch.

The solution of the size/homogeneity/horizon problem is found if the
presence of the large homogenous patch $l_{H_0}$ with Planck energy
density is explained. One way to achieve this is to blow up an
initial patch of size $l_x = \Lambda^{-1} = 10^{-x} \, GeV^{-1}$ to
size $l_{H_0}$ without significant loss of energy density. A period
of accelerated expansion of the universe - inflation~\cite{Guth} is
usually introduced for this purpose. Problems of inflationary
cosmology~\cite{Brandenberger:1999sw} motivate attempts to find
alternative explanations.

The paper~\cite{size} pretends to answer the
size/homogeneity/horizon problem by a long period of intermediate
evolution in a $d+4$ extra-dimensional universe. This allows slow
enough decrease of the initial energy density without going into an
inflationary regime. The scenario of the extra dimensional universe
which is developed in the paper~\cite{size} has similar initial
conditions to which is assumed in the hot Big Bang Cosmology, the
only difference is number of dimensions. Namely, existence of a
higher dimensional universe of linear size ($\Lambda^{-1}$) and
energy density $\rho \sim \Lambda^{4+d}$ is assumed. To follow the
evolution of the universe we assume that its content can be
described by a perfect fluid and a weak anisotropic potential $V$.
Initially, the evolution is governed by the perfect fluid (bulk
matter). Isotropic expansion ends when $V$ cannot be neglected any
more. Then anisotropic evolution begins. The ultimate requirement on
$V$ is to prevent the decompactification of the extra dimensions.
Once the energy density is dominated by the potential, the extra
dimensions shrink while our dimensions continue to expand. Late time
four dimensional cosmology is recovered once the extra dimensions
are stabilized.

In this paper we show that the minimal value for $\Lambda$ is
$10^{25} \, GeV$. The analysis is based on the fact that the
universe has to achieve the current size ($H_0^{-1}$). General
arguments of the Section II show that 3+1 dimensional power law
expanding universes with the one scale initial conditions (universe
of linear size ($\Lambda^{-1}$) and energy density $\rho \sim
\Lambda^{4}$) have to have $\Lambda \geq 10^{25} \, GeV$ if no
accelerated expansion is assumed, and for all $\Lambda$, a radiation
dominated epoch alone leaves the size/homogeneity/horizon problem
unresolved. Note, in general the radiation dominated epoch can
explain the size/homogeneity/horizon problem if the initial energy
density is assumed to be sufficiently high and there is no demand of
the one scale initial conditions~\cite{Kofman:2002cj}.
In the Section III, the scenario discussed in~\cite{size} is
described from the effective four dimensional point of view and
potentially can fit other extra-dimensional
constructions.~\footnote{In the original paper~\cite{size}, it
remains unclear which of the scale factors, $a(t)$ - the 3
dimensional in the $d+4$ dimensions or $\alpha(t)$ - the scale
factor in the effective four dimensional setup, is supposed to grow
up the required amount. See~\cite{thesis} for the correction.}
Hence, we do not inter into details of the stabilization stage. We
find analytical approximations for the evolution of the scale factor
and the extra dimensions. In particular, we find a scaling solution
which is in effect when two fluid components (bulk matter and a
potential) cannot be neglected. The new result shows that whether
expansion is inflationary depends not only on the slope of the
potential but also on the number of the extra dimensions (d) and the
equation of state of the bulk fluid ($w$). We provide a numerical
example for demonstration purposes.

\section{General Arguments}

Let us start with an isotropic patch of the size $\Lambda^{-1} =
10^{-x} \, GeV^{-1}$ and an initial energy density of order
$\Lambda^4$. These initial conditions are referred in the following
as one scale initial conditions. To reach the size of the currently
visible patch, $H^{-1}_0 \approx 10^{42} \, GeV^{-1}$, it has to
expand by $x+42$ orders of magnitude. The evolution of the maximally
symmetric hyper-surface without cosmological constant is described
by
Friedmann-Robertson-Walker (FRW) equations \beqa 3 m^2_{p} H_E^2 &=& \tilde{\rho} \\
-2 m^2_{p} \dot{H}_E &=& \tilde{\rho} + \tilde{p} \, , \label{FRW}
\eeqa where $m_p= 2.43 \, { \mbox x } \, 10^{18}$ GeV is the reduced
Planck mass. The notations are as follows $\tilde{\rho}$ is the
energy density, $\tilde{p}$ is the pressure, $\alpha$ is the scale
factor and $H_E = \dot \alpha/\alpha$ is the Hubble parameter in the
Einstein frame. In a power law expanding universe $ \alpha \propto
t^k \, $ with $k>0$, $ H_E \propto t^{-1} \, , \, \tilde{\rho}
\propto t^{-2} $ and, $\tilde{\rho} \propto \tilde{p}$. At the time
of the Big Bang Nucleosynthesis ($\tilde{\rho}_r \approx MeV^4$) and
afterwards the content of the universe is known to a high precision
and so is the value of $k$. See e.g.~\cite{Simha:2008zj}. During
this period the universe has expanded by $11-12$ orders of
magnitude. Up to the time of the Big Bang Nucleosynthesys \beq
\frac{\alpha_r}{\alpha_i} = \Big(
\frac{\tilde{\rho_i}}{\tilde{\rho_r}} \Big)^{k/2} \, , \eeq  where i
stands for initial and r - for the final values. For an arbitrary
$k$, $\alpha$ grows by a factor of $10^{2k \, (x+3)}$ \beq
\frac{\alpha_r}{\alpha_i} = \Big( \frac{10^{x} GeV}{10^{-3} GeV}
\Big)^{2k} = 10^{2k \, (x+3)} \, . \eeq The requirement that the
scale factor grows $x+31$ orders of magnitude imposes $x
> 25$ for $k < 1$. Therefore, in a power law expanding universe, a
non-inflationary solution of the size/homogeneity/horizon problem
requires the scale of initial energy density to be larger than
$\Lambda = 10^{25} \, GeV$. As this scale is much larger than the
Planck scale, the FRW equations~(\ref{FRW}) can no longer be
trusted. Note that in a radiation dominated universe the solution of
the problem is technically impossible. Explicitly, in the radiation
dominated universe ($k=1/2$) the available $x+3$ orders of magnitude
are much smaller than the required $x+31$ orders of magnitude.
Hence, a new stage of evolution with a different equation of state
is required. The arguments above are qualitatively different from
the standard arguments like in~\cite{Kofman:2002cj} where there is
no demand of the one scale initial conditions. In particular, a
consequence of no demand of the one scale initial conditions allows
to explain the size/homogeneity/horizon problem by the radiation
dominated epoch alone.

\section{Calculations in the Einstein Frame}

We assume that the evolution of the $d+4$ dimensional Universe is
governed by the Einstein equations. Let $G_{ab}$ be the metric for
the full space-time with coordinates $X^a$. The line element of the
spatially flat but anisotropic universe is
\begin{equation}
ds^2  =  G_{ab} dX^a dX^b  \, = \, - dt^2 + a(t)^2 d{\bf x}^2 +
b(t)^2 d{\bf y}^2 \, ,
\end{equation}
where ${\bf x}$ denotes the three coordinates parallel to our
visible three dimensional Universe and ${\bf y}$ denotes the
coordinates of the $d$ perpendicular directions. The action of the
Universe is described by
\begin{equation}
S  =  \int d^{d+4}X \sqrt{-\det G_{ab}} \left\{ \frac{1}{16 \pi
G_{d+4}} R_{d+4} + \hat{\cal L}_M \right\} \,,
\end{equation}
where $R_{d+4}$ is the $d+4$ dimensional Ricci scalar and $\hat{\cal
L}_M$ is the matter Lagrangian density with the metric determinant
factored out.

To follow the evolutions of 'our' three spatial dimensions from the
effective four dimensional point of view, we replace the $b(t)$ by a
canonically normalized scalar field $\varphi(t)$ which is related to
$b(t)$ through
\begin{equation}
\label{rel} \varphi  = \beta^{-1} m_{p} \ln (b) \, ,
\end{equation}
where we have introduced
\begin{equation}
 \beta^{-1} \equiv
{\sqrt{d(d+2) \over 2}} \,.
\end{equation}
To justify the use of the effective four dimensional treatment we
need to make sure that 'our' three dimensions are large comparing to
the extra dimensions. We also assume that all dimensions in the
universe besides our 3 start small, of size $\l = \Lambda^{-1}$,
where $\Lambda$ is the initial scale of the energy density.
Furthermore, we assume that there is a patch in our three large
dimensions of size $\l = \Lambda^{-1}$ with the same energy density,
and from now on we concentrate on the evolution of this patch. In
terms of $\varphi$ the effective reduced four dimensional action
after performing a conformal transformation (with the help
of~\cite{Carneiro:2004rt}) to arrive at the Einstein frame is
\begin{eqnarray}
S  =  \int d^4 x \sqrt{- \det \tilde{g}_{\mu \nu}} \left\{
\frac{1}{2} m^2_{p} R_4 \right. &-& \left. \frac{1}{2}
\tilde{g}^{\mu\nu} \tilde{\nabla}_\mu \varphi \tilde{\nabla}_\nu
\varphi \right. \\ &+& \left. {\cal V} e^{- d \varphi/m_{p}\beta}
\hat{\cal L}_M \right\}\,, \nonumber
\end{eqnarray}
where
\begin{equation}
{\cal V} \, = \, \int d^d{\bf y} \, = \, l^d
\end{equation}
is the coordinate volume of the extra dimensions, and
$\tilde{g}_{\mu \nu}$ is the 4d metric in the Einstein frame, \beq
ds_E^2 \, = \, \tilde{g}_{\mu \nu} dx^{\mu} dx^{\nu}  \, = \, - dt^2
+ \alpha(t)^2 d{\bf x}^2 \, . \eeq The matter content (bulk matter
and a potential) dictates the following Lagrangian~\footnote{This is
the same Lagrangian as assumed in the paper~\cite{size}.} \beq
\hat{\mathcal{L}}_M = -\rho - V(b) \eeq where $\rho$ is the energy
density of the bulk matter and $V$ is a potential responsible for
the contraction of the extra dimensions. Potential $V$ arises as a
result of the construction of the space-time manifold. Specifically,
in the paper~\cite{size}, $V$ is a confining potential between the
orbifold fixed planes and has power law dependence on the scale
factor $b$.

It is widely used that the part of the Lagrangian  \beq
\hat{\mathcal{L}}_{M1} = -\rho \eeq provides the energy-momentum
tensor for a perfect fluid, see e.
g.~\cite{Mukhanov:1990me,Battefeld:2004xw}, while a derivation of
the energy momentum from the action is left to a reader. In this
paragraph we show the derivation. The energy-momentum tensor is
obtained from variation of the action with respect to the metric
\beq T_{a b} = -2 \frac{\delta \hat{\mathcal{L}}_{M1}}{\delta G^{a
b}} + G_{a b} \hat{\mathcal{L}}_{M1} \, . \label{highenmo} \eeq We
can replace variation with respect to the metric with variation with
respect to the determinant of the metric. \beq \delta \sqrt{-\det
G_{ab}} = - \frac{1}{2} \sqrt{-\det G_{ab}} G_{ab} \delta G_{ab} \,
. \eeq In the case $\delta G_{ab} \neq 0$, \beq \frac{\delta
\hat{\mathcal{L}}_{M1}}{\delta G^{a b}} = -\frac{1}{2} \sqrt{-\det
G_{a b}} G_{a b} \frac{\delta \hat{\mathcal{L}}_{M1}}{\delta
\sqrt{-\det G_{a b}}} \, . \eeq However, the zeroth component of the
metric is constant and, hence, $\delta G_{0 0} =0$. To proceed let
us define $U_{a}$ to be the unit timelike vector orthogonal to the
spatial slices, so \beq G_{ab} = -U_a U_b + \gamma_{a b}
\label{relationG} \eeq with $U^a \gamma_{a b} = 0$. Therefore, \beq
\frac{\delta \hat{\mathcal{L}}_{M1}}{\delta G^{a b}} = -\frac{1}{2}
\sqrt{-\det G_{a b}} \gamma_{a b} \frac{\delta
\hat{\mathcal{L}}_{M1}}{\delta \sqrt{-\det G_{a b}}} \, . \eeq The
higher dimensional energy-momentum tensor for a perfect fluid is
supposed to take the familiar form \beq T_{a b} = (1+w)\rho U_a U_b
+ w \rho G_{a b} \, , \label{famformEMT} \eeq where $w$ is an
equation of state parameter of the bulk matter. Moreover, we treat
the case when $w$ is a constant. The conservation of the higher
dimensional energy-momentum tensor for a perfect fluid dictates \beq
\rho \propto a^{-3(1+w)} \, b^{-d(1+w)} \, . \label{rho} \eeq
Variation of~(\ref{rho}) with respect to $ \sqrt{-\det G_{a b}} =
a^3 b^d$ is equal to $-(1+w)\rho / (\sqrt{-\det G_{a b}})$. Plugging
back all the above into~(\ref{highenmo}), we obtain \beq T_{a b} =
-\rho G_{a b} + (1+w) \rho \gamma_{a b} \, . \eeq The familiar form
of the energy momentum tensor for a perfect fluid~(\ref{famformEMT})
is recovered making use of the relation~(\ref{relationG}).

The effective 4 dimension Lagrangian is \beq \tilde{\cal L} = -
\frac{1}{2} \tilde{g}^{\mu\nu} \tilde{\nabla}_\mu \varphi
\tilde{\nabla}_\nu \varphi \, - \Omega - V(\varphi) \, \eeq
with \beqa  \Omega &\equiv& {\cal V} \, \rho \, e^{- d \beta \varphi/m_{p}}
= \Omega_0 \alpha^{-3(1+w)}e^{-\gamma_2 \varphi} \, \label{Omega2} \\
V(\varphi) &\equiv& {\cal V} \, V(b) \, e^{- d \beta \varphi/m_{p}}
\, , \eeqa where \beq
      \gamma_2 = \frac{d(1-w) \beta}{2 m_p} \, .
\eeq In the above, $\Omega_0$ is a constant and we made use of
$\alpha = a b^{d/2}$.

The 4 dimensional energy-momentum tensor takes the form \beq
\tilde{T}_{\mu \nu} = \tilde{\nabla}_\mu \varphi \tilde{\nabla}_\nu
\varphi - \frac{1}{2} \tilde{g}_{\mu \nu} \tilde{g}^{\rho \sigma}
(\tilde{\nabla}_\rho \varphi )(\tilde{\nabla}_\sigma \varphi) +
(1+w) \Omega \tilde{\gamma}^{\mu \nu} - (\Omega +
V(\varphi))\tilde{g}_{\mu \nu} \, , \eeq where \beq \tilde{g}_{\mu
\nu} = -\tilde{U}_\mu \tilde{U}_\nu + \tilde{\gamma}_{\mu \nu} \, .
\label{relationg} \eeq The equation of motion for the field
$\varphi$ is \beq \ddot{\varphi} + 3 H_E \dot{\varphi} + \Omega'
 + V'(\varphi) = 0 \, \label{eqmot1} \, ,\eeq
where $'$ denotes differentiation with respect to $\varphi$.
Equation~(\ref{eqmot1}) together with the equation for the
conservation of the energy momentum tensor \beq \tilde{\nabla}^\mu
T_{\mu\nu}=\dot{\varphi} \ddot{\varphi} + V'(\varphi) \dot{\varphi}
+ \dot{\Omega} + 3 H_E \dot{\varphi}^2 + 3 H_E (1+w) \Omega = 0 \,
\label{consenmom} \eeq
is consistent if \beq \Omega = \Omega(\varphi) \alpha^{-3(1+w)} \, ,
\label{Omega} \eeq
which is indeed the case (see~(\ref{rho})). In the Einstein frame
the energy density is $\tilde{\rho} = \tilde{T}_{00}$ and momentum
is $\tilde{p}=\tilde{T}_{ii}/\tilde{g}_{ii}$ with $i=1,2 \mbox{ or }
3$. Then, the FRW equations read: \beqa 3 m^2_{p} H_E^2 &=&
\tilde{T}_{00} = \left\{\frac{1}{2}
\dot{\varphi}^2 + V(\varphi) + \Omega \right\} \label{TOO} \\
-2 m^2_{p} \dot{H}_E &=&\tilde{T}_{00} + \tilde{T}_{11}/\alpha^2 =
\left\{ \dot{\varphi}^2 + (1+w) \, \Omega\right\} \, . \label{dotHe}
\eeqa  The three equations~(\ref{Omega2},\ref{TOO},\ref{dotHe})
fully determine the evolution of our three dimensions. Note that if
the potential is positive the scale factor of our four dimensions,
$\alpha$, always grows since $H_E^2$ never reaches zero, hence,
there is no collapse in the effective 4d Einstein frame. The
effective equation of state in the Einstein frame for $-1<w<1$ is
\beq 1 \geq \frac{\tilde{p}}{\tilde{\rho}} = \frac{1/2 \,
\dot{\varphi}^2 - V(\varphi) + w \, \Omega}{1/2 \, \dot{\varphi}^2 +
V(\varphi) + \Omega} \geq -1 \, . \eeq

Two forms for the potential have been considered in the literature:
$V(\varphi) = M_0 e^{\gamma_1 \varphi}$ in~\cite{size} and
$V(\varphi) = V_0 (1- \zeta e^{-\gamma_1 \varphi})$
in~\cite{Shuhmaher:2005pw}. The collapse of extra dimensions is
possible only if $\gamma_1 > 0$. In the above $M_0$, $V_0$, $\zeta$
and $\gamma_1$ are constants which depend on the initial conditions.
The second case $V(\varphi) = V_0 (1- \zeta e^{-\gamma_1 \varphi})$
leads to inflation and is extensively treated
in~\cite{Battefeld:2006cn}. This is the case when inflation is free
from the problem of initial conditions. In this paper, we
concentrate on the case $V(\varphi) = M_0 e^{\gamma_1 \varphi}$
which has not been treated fully. In the following, we include the
residual energy density of the bulk matter, the detail which has
been neglected in~\cite{size}, and calculate the evolution of the
scale factor.

Let us divide the solution to different regions. In some regions the
solutions for $\varphi$ and $\alpha$ can be approximated by
\beqa
\varphi &=& A_1 + A_2 \ln(t) \\
\alpha &=& A_3 t^{k} \, . \eeqa

where $A_1$, $A_2$, $A_3$ and $k$ are constants. Initially, when
$\Omega
> V$ the solution is

\beqa A_2 &=& \frac{2 m^2_p \gamma^2_2 + 3 (1+w)^2 -
6(1+w)}{2\gamma_2 (m^2_p \gamma^2_2 - \frac{3}{2}(1+w)^2 - 3 (1+w))}
\\ &\pm& \frac{\sqrt{2 m^2_p \gamma^2_2 - 3(1+w)^2 - 16 m^2_p
\gamma^2_2 (m^2_p \gamma^2_2 - \frac{3}{2}(1+w)^2 -
3(1+w))}}{2\gamma_2 (m^2_p \gamma^2_2 -
\frac{3}{2}(1+w)^2 - 3 (1+w))} \\
k &=& \frac{2-\gamma_2 A_2}{3(1+w)} \\
A_2 &-& 3 k A_2 = \gamma_2 \Omega_0 A^{-3(1+w)}_3 e^{-\gamma_2 A_1}
\, , \eeqa

where $A_1$ and $A_3$ are determined by initial conditions and fit
to later evolution. If $\dot{\varphi}$ dominates the evolution of
the universe, the solution is the one of shift matter,

\beqa
k &=& 1/3 \\
A_2 &=& \sqrt{\frac{2}{3}} m_p \eeqa

Later on when the potential cannot be neglected any more, the
description of the evolution deviates from the provided
in~\cite{size}. Once the extra dimensions shrink, the solution
becomes \beqa
k &=& \frac{2}{3} \frac{\gamma_1 + \gamma_2}{\gamma_1 (1+w)} \\
e^{\gamma_1 A_1} &=& \frac{3 m^2_p k^2 \gamma_2 -\frac{1}{2}
A^2_2 \gamma_2 + A_2 + 3 k A_2}{M_0 (\gamma_1 + \gamma_2)} \\
A_2&=& - \frac{2}{\gamma_1} \, . \eeqa This is a scaling solution
since the potential and the residual energy density in $\Omega$
dilute at the same rate. The solution is non-accelerating if $k<1$.
To see the difference from the conclusions of paper~\cite{size},
recall that the solution, where only the potential is taken into
account, is non-accelerating if $\gamma_1 \geq
\sqrt{2}/m_p$~\cite{thesis} (the slow roll parameter
$\epsilon=m_p^2/2 \, (V'/V)^2 > 1$). Namely, $k$ was determined
entirely by $\gamma_1$. In the present full analysis, we see that
also $\gamma_2$ and $w$, i.e. properties related to the bulk fluid,
are relevant for the determination of $k$. This difference allows
one to find examples where non-inflationary potentials lead to
inflation.

\begin{figure}[tb]
  \includegraphics[width=0.5\textwidth, angle =-90]{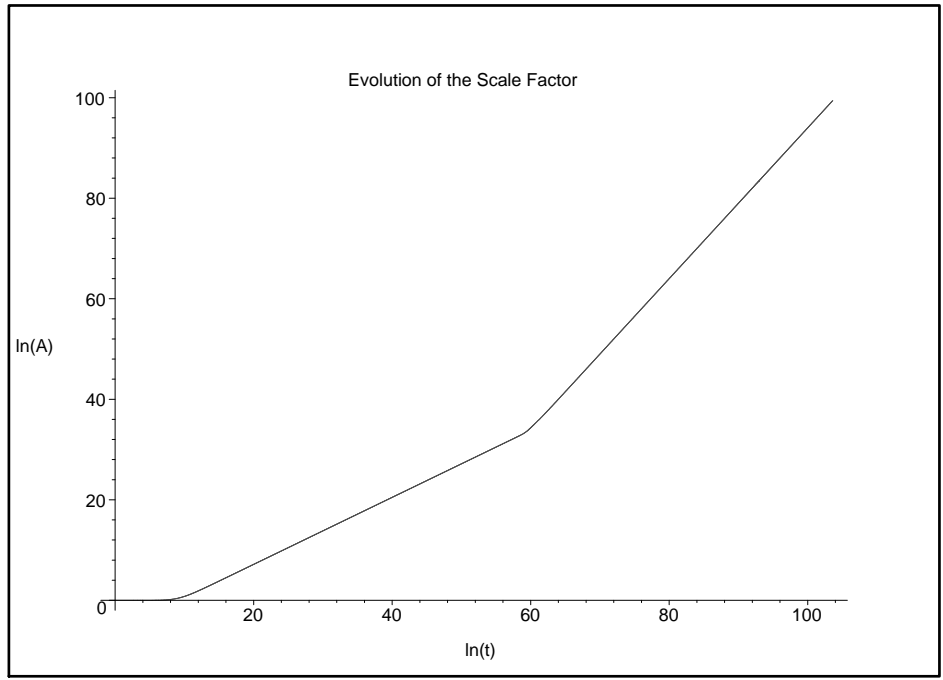}
 \caption{\label{Fig1}
The graph illustrates logarithm of the scale factor ($ln(A)$) versus
logarithm of time. While the extra dimensions expand (phase 1), our
three dimensions decelerate $\ln(\alpha(t)) = 0.66 \, \ln(t) - 6$
and $k=0.66$. Once the extra dimensions contract (phase 2), our
dimensions accelerate $\ln(\alpha(t)) = 1.5 \, \ln(t) - 55$, namely
$k = 1.5$ as expected from the scaling solution. }
\end{figure}

\begin{figure}[tb]
  \includegraphics[width=0.5\textwidth, angle=-90]{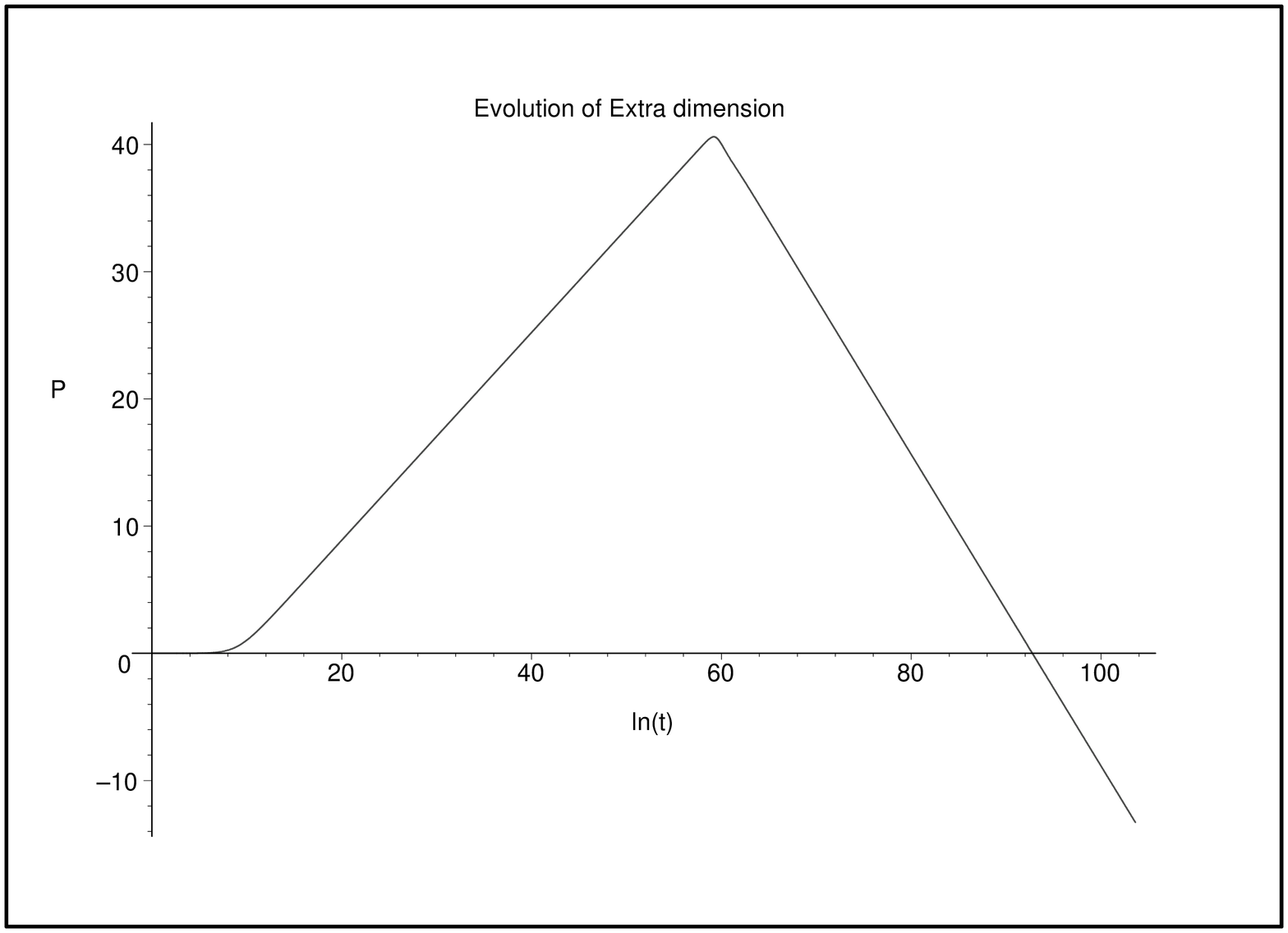}
 \caption{\label{Fig2}
The graph illustrates evolution of $\varphi$ (P)- the size of the
extra dimensions as function of time. Initially $\varphi$ growing
and then decreasing. The two phases correspond to the expansion and
contraction phases of extra dimensions.}
\end{figure}

The following example demonstrates the above statement. Let us take
the initial value for $\alpha_{in}=ab^{d/2}= 1$ and choose
$\varphi_{in} = 0$. Further, we choose $\Lambda = 10^{-2} \, m_p$
and find $\Omega_0 = \Lambda^4 = 10^{-8} \, m^4_p$. The symmetry in
the initial conditions allows to assume that $\dot{a}/a$ and
$\dot{b}/b$ are of the same order. As a consequence, $H_E^2$ and
$\dot{\varphi}^2$ in the equation~(\ref{TOO}) are of the same order.
This permits us to estimate the initial value for
$\dot{\varphi}(t_i) = 10^{-4} \, m^2_p $. Taking the scale of the
potential $V(\varphi)$ to be $10^{-5} m_p$, we get $M_0=10^{-80}\,
m_p^4$ and, hence, at the end of the contraction phase of the extra
dimensions, the scale of the potential energy is higher the scale of
the Big Bang Nucleosynthesis - 10 MeV. The choice of $d=6$ and
$w=-1/3$ gives $\gamma_2 = \sqrt{2/3}/m_p$. The choice of $\gamma_1
= \sqrt{8/3}/m_p > \sqrt{2}/m_p$ leads to a non-inflationary
solution if only the potential is taken into account as
in~\cite{size}. However, as noticed above, the scaling solution
depends also on the values of $\gamma_2$ and $w$. Hence, as the
calculation shows if the bulk fluid is taken into account ($k=1.5$),
there is acceleration. The numerical solution for the scale factor
(see Fig.~\ref{Fig1}) supports this conclusion. To demonstrate the
evolution of the extra dimensions, we solve numerically for
$\varphi$ (see Fig.~\ref{Fig2}).

\section{Conclusions}

This paper discusses the entropy and size/homogeneity/horizon
problems in power low expanding universes. We show that the
radiation dominated stage alone in the universe with the one scale
initial conditions cannot explain the issue regardless of the
initial scale. A non-inflationary solution of the problems in the
initially one scale universe is achieved only if this scale is
allowed to be higher than $10^{25}$ GeV. At this energy scale,
corrections to the FRW equations~(\ref{FRW}) should apply which are
the subject of a followup work.

The extra-dimensional scenario developed in~\cite{size} is
reconsidered in Section III. We perform the analysis of the
evolution of a $d+4$ dimensional universe in the effective Einstein
frame. Since $H_E$ is always positive and decreasing, the universe
in the 4d Einstein frame never undergoes the stage of contraction.
The earlier paper~\cite{size} neglects the bulk energy density in
the investigation of the second stage of the evolution of the
universe, the stage when the potential responsible for the
contraction of the extra dimensions becomes important. Our new
result is that the question whether the expansion is inflationary
also involves the behavior of the bulk fluid. The numerical example
explores the case where the potential itself does not cause
inflation but the inclusion of the bulk energy density (the scaling
solution) changes this.

The scenario of Section III basically describes a power law
expanding universe. Throughout the evolution of the universe the
power of the scale factor ($k$) changes. As shown in Section II, the
minimal scale to resolve the size/homogeneity/horizon problem in
power law expanding universe without inflation is $\Lambda = 10^{25}
GeV$. It is not generic at all, that such or higher energy density
scale is allowed in a particular string theory model. In addition,
the low energy field theory constructions should not be trusted at
this or above scale since stringy effects become important. In this
light the validity of the string theory based
construction~\cite{size} to solve the problem is put under question.

Difficulty to find a non-inflationary solution based on this
scenario might lead to interesting models of inflation (like
in~\cite{Shuhmaher:2005pw} and~\cite{Battefeld:2006cn}). An
interesting feature of the setup discussed here is the freedom from
the problem of initial conditions on the onset of inflation, namely,
no fine tuning of the slope of the potential. One starts with a
dense unique scale patch filled with bulk matter and ends up with
inflation. Note that in the cases~\cite{Shuhmaher:2005pw}
and~\cite{Battefeld:2006cn}, the energy density in the bulk matter
during the second stage is rapidly diluted and does not have much
influence on the behavior of the scale factor.

\begin{acknowledgments} It is a pleasure to thank Ruth Durrer and Stefano Foffa for
proofreading of the manuscript and comments. This work is supported
by the Tomalla Foundation.
\end{acknowledgments}

\end{document}